# Effect of mounting strut and cavitator shape on the ventilation demand for ventilated supercavitation

Siyao Shao[1,2], Arun Balakrishna[1,2], Kyungduck Yoon[1,2], Jiaqi Li[1,2], Yun Liu[3], Jiarong Hong[1,2, *]

1. Saint Anthony Falls Laboratory, 2 3rd Ave SE, University of Minnesota, Minneapolis, MN, 55414, USA
2. Department of Mechanical Engineering, University of Minnesota, Minneapolis, MN, 55414, USA
3. Department of Mechanical and Civil Engineering, Purdue University Northwest, Westville, IN 46391, USA

* Email address of the corresponding author: jhong@umn.edu

## Abstract

The present work systematically investigates the effect of cavitator mounting strut and cavitator shape on the ventilation demand to form and sustain a ventilated supercavity under different flow conditions. Three cavitators of different shapes (i.e. 2D cavitators including triangle and disk shape, and 3D cone-shaped cavitator) with the same frontal area are employed for the experiments. The cavitator is connected to a ventilation pipe extended upstream from a mounting strut, referred to as forward-facing model (FFM). The minimal ventilation coefficients to generate ($C_{Qf}$) and to sustain ($C_{Qc}$) a supercavity are measured over a wide range of Froude number ($Fr$) for each cavitator. Images of overall cavity shapes and topology near closure region as well as the cavity pressure under different experimental conditions are captured simultaneously. The results are compared across different cavitator shapes and the disk cavitator situated on the downstream side of the mount strut, referred to as backward-facing model (BFM). Similar $C_{Qf} - Fr$ curves were observed in BFM and FFM-configured disk cavitators. The cone-shaped cavitator requires the least amount of ventilation to generate a supercavity among all different shapes across the range of $Fr$ in our experiments. The $C_{Qc}$ of disk FFM is lower than that of its BFM counterpart at small $Fr$ and exceeds the BFM $C_{Qc}$ with further increase of $Fr$. The cone cavitator has the smallest $C_{Qc}$ among all the cavitators across the range of $Fr$ in our experiments. To elucidate the trends of $C_{Qc}$ upon changing $Fr$ and cavitator shape, the geometry of supercavity under $C_{Qc}$ including its overall shape, the cavity maximum diameter ($D_{max}$) and half length ($L_{1/2}$) are also investigated. Both $D_{max}$ and $L_{1/2}$ show an increasing then plateauing trend upon increasing $Fr$ across different FFM cavitators despite the smaller values for the 3D cone cavitator. Subsequently, such cavity geometric information and cavity pressure measurements in conjunction with high speed imaging of re-entrant jet are employed to estimate the re-entrant jet momentum under different $Fr$ for disk and cone cavitators. The estimated re-entrant jet momentum shows reasonable match with the ventilation air momentum under $C_{Qc}$ in lower $Fr$ for both cavitator cases, with the disk cavitator case yielding significantly stronger re-entrant jet, providing support to the re-entrant jet mechanism governing on the cavity collapse. Our study sheds some light on the cavitator design and ventilation strategy for a supercavitating vehicle in practice.

# 1. Introduction

Ventilated supercavitation, i.e. a special case of cavitation in which the cavitating body can be enclosed in a gas bubble generated by injecting gas behind a cavitator, has gained substantial attention for its potential capabilities in high speed underwater applications [1]. Traditionally, such phenomenon can be characterized by using non-dimensionalized parameters such as ventilated cavitation number, $\sigma_C = 2(P_\infty - P_c)/(\rho_W U^2)$, Froude number, $Fr = U/\sqrt{gd_c}$, and ventilation coefficient, $C_Q = \dot{Q}/(Ud_c^2)$, where $P_\infty$ and $P_c$ refer to the test-section pressure upstream of the cavitator and the cavity pressure, while $\rho_W$, $U$ and $g$ correspond to liquid density, the free stream velocity in the test-section and gravitational acceleration, respectively. In the definition of $Fr$ and $C_Q$, $d_c$ denotes the cavitator diameter and $\dot{Q}$ is the volumetric air flow rate.

The operation of a ventilated supercavitating vehicle depends on its ability to supply sufficient gas to fill the cavity at different flow conditions and different stages over the whole course of the vehicle operation [2]. Therefore, the design of a ventilated supercavitating vehicle should consider the ventilation demand of the supercavity especially during its formation and sustenance to achieve an optimal ventilation gas storage and an extension of the range of the vehicle cruising. Recently, a number of studies have investigated the ventilation demands of ventilated supercavity under a wide range of flow conditions for disk cavitators [2-5]. Specifically, Kawakami and Arndt [3] reported ventilated hysteresis which refers to a phenomenon that the ventilation demand to sustain a cavity is substantially lower than the requirement to form a cavity. Therefore, a design strategy diligently considering ventilation hysteresis can substantially decrease the total gas storage of a ventilated supercavitating vehicle. Karn et al. [2] conducted a measurement of ventilation demand of the cavity generated using varying sizes disk cavitators over a wide range of flow condition. The results demonstrated that the ventilation demand for cavity formation is much greater than that requires to sustain a supercavity, and the formation ventilation coefficient ($C_{Qf}$) displays an increasing then decreasing trend over increasing $Fr$. They suggested that supercavity formation is correlated with the coalescence efficiency of the small bubbles generated at a given flow condition. Such hypothesis has been further supported by a comparison of the $C_{Qf}$ from different facilities in Shao et al. [5]. They attributed the discrepancy in the $C_{Qf}$ from different facilities to the influence of the mismatched Reynolds number on the bubble coalescence process. On the other hand, as shown in Karn et al. [2], the minimal ventilation coefficient to sustain a supercavity, i.e. the ventilation threshold below which the supercavity collapses (referred to as cavity collapse ventilation coefficient $C_{Qc}$), exhibits a monotonic decreasing then plateauing trend with increasing $Fr$. Karn et al. [2] noted that this trend may be caused by the variation of non-dimensionalized pressure difference across the cavity closure according to the general framework presented in [4]. However, despite the abovementioned studies, systematic investigation of $C_{Qf}$ and $C_{Qc}$ has not yet been conducted under different cavitator mounting struts and cavitator shapes. The knowledge derived from such investigation could provide useful insight for optimizing the cavitator design of a supercavitating vehicle to minimize its gas storage.

In general, the mounting strut of a cavitator has two types of configuration, i.e. backward-facing model (BFM) and forward-facing model (FFM) [6]. In BFM, a cavitator is supported by a thin hydrofoil situated at the front of cavitator with a ventilation pipe enclosed inside the hydrofoil. Such configuration can generate a "free-standing cavity" which does not enclose a solid object nor interact with any solid surfaces. The FFM, however, has a ventilation pipe behind the cavitator for both gas supply and cavitator support. This configuration resembles the case of a ventilated supercavitating vehicle in practice, in which the supercavity is in contact with the solid surface of the vehicle. Previous investigations have mainly studied the effect of mounting strut on cavity geometry, cavity pressure and the force exerted on the cavity [3, 6-8], but very few considers its effect on the ventilation demand of a cavity.

As for cavitator shape, the prior studies have been focused on its impact on drag reduction efficiency and cavity geometry. Specifically, Semenenko [9] derived the semi-empirical formulas of drag coefficients for disk and cone-shaped cavitators based on experimental data. Combining numerical simulation and experimental measurements, Ahn et al. investigated the change of cavity dimension (i.e., length and diameter) and cavitation number for wedge and cone cavitators under different ventilation and flow conditions [10]. Specifically, they showed that cone and wedge cavitators result in a shorter and slimmer cavity compared to that generated by a disk cavitator with the same frontal area from [11]. Recently, Moghimi et al. [12] conducted experiments using disk, cone and parabolic cavitators, and compared the corresponding cavity dimension, cavitation number and the overall drag reduction effect, which shows the parabolic cavitator leads to minimal drag among the three cavitators of the same frontal area. Although the abovementioned investigations have provided important guidance for cavitator design, there is still a dearth in the study of the effects of cavitator shapes on $C_{Qf}$ and $C_{Qc}$.

Based on the above literature review, our study focuses on a systematic experimental investigation of the effect of mounting strut and cavitator shape on the $C_{Qf}$ and $C_{Qc}$ of ventilated supercavitation. The paper is structured as follows: Section 2 describes the experimental method including the flow facility, the cavitator design and the measurement setup. Section 3 presents the experimental results and the corresponding analysis, which is followed by a conclusion and discussion in Section 4.

## 2. Experimental Setup and Methodology

The experiments are conducted at the high-speed water tunnel (Fig. 1) in the Saint Anthony Falls Laboratory (SAFL). The water tunnel has a test section of 1200 mm (length) × 190 mm (width) × 190 mm (height), and is capable of operating at flow speed up to 20 m/s with a turbulence level of 0.3 %. A large dome-shaped settling chamber situated upstream of the test section has the capability of fast removal of gas bubbles which allows a continuous operation of cavitation and ventilation experiments. In the recent years, this facility has been used for a number of supercavitation [2-5] and hydrofoil aeration experiments [13].

Two 2D-shaped cavitators (i.e., triangle and disk) and one 3D cone-shaped cavitator are fabricated using a Lulzbot Taz 6 3D printer (Fig. 2). The polylactic acid (PLA) filament is chosen as the material of the cavitators since the PLA filament has anti-water corrosion and leakage-resistance features and the PLA filament can achieve a high-resolution printing corresponding to a layer height of 0.25 mm. The cavitators have the same frontal areas as the 30-mm-diameter disk shape and the cone cavitator has a draft angle $\theta = 31°$. The cavitators are mounted on a forward-facing model (FFM) as shown in Fig. 3. Note that the cavitator diameter (i.e., $d_c$) is set to be 30.0 mm for calculating non-dimensionalized parameters in the current study.

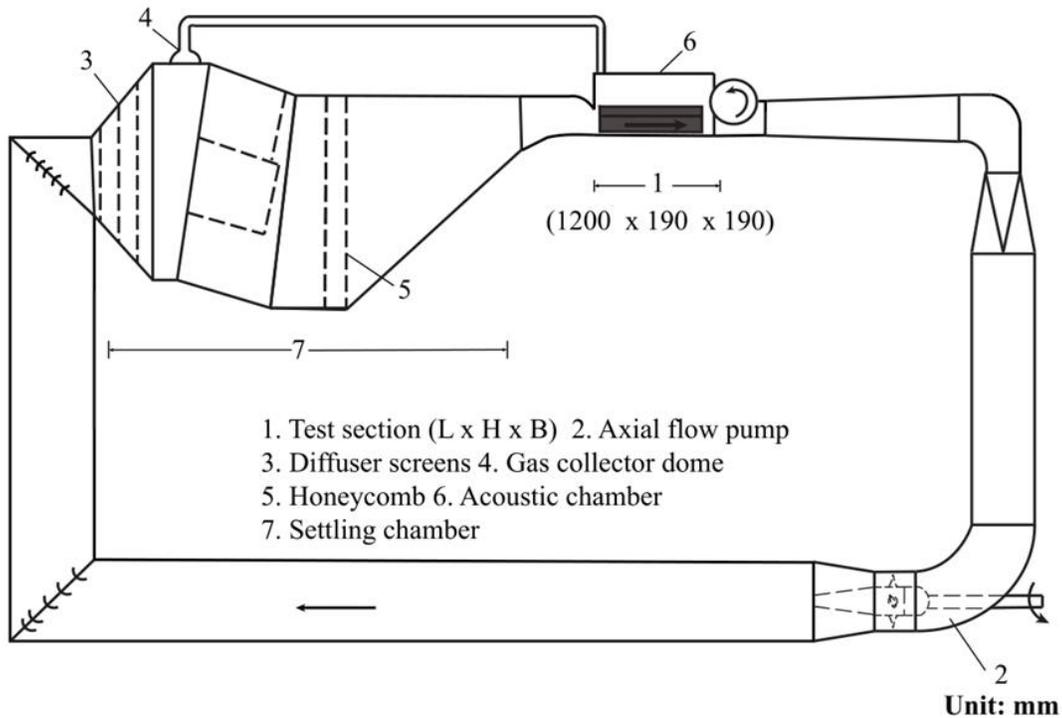

Figure 1: Saint Anthony Falls Laboratory (SAFL) cavitation water tunnel used for the current experiments. This schematic is adapted from [5].

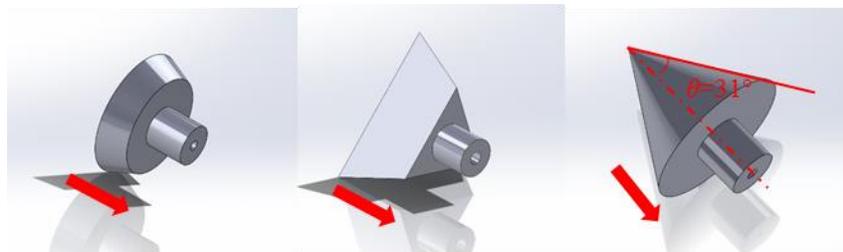

Figure 2: 3D Computer renderings of the disk, triangle, and cone cavitators used in the current experiments, respectively. Note that the draft angle for the cone cavitator is 31°. Arrows designate the flow direction.

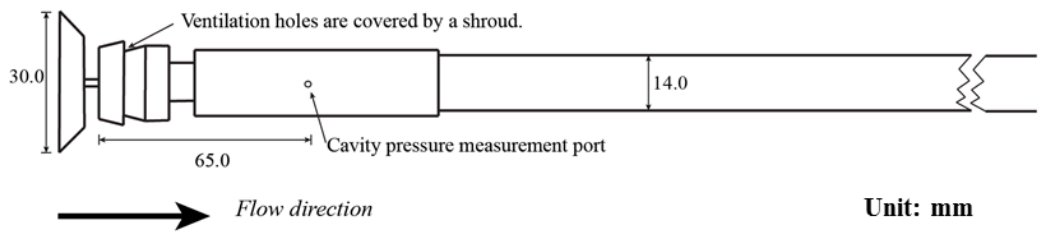

Figure 3: Forward-facing model (FFM) used in the experiments. Note that the disk cavitator is shown in this schematic. This schematic is adapted from [5].

In the experiments, the $C_{Qf}$ and $C_{Qc}$ of three cavitators are examined over $Fr$ from 5.6 to 18.4 (corresponding to a flow speed from 3 m/s to 10 m/s). The air flow rate is regulated by a FMA-2609A mass flow controller with a unit of standard liter per minute (SLPM), which is the volumetric flow rate at the standard temperature (273.15 K or 0 °C) and standard pressure (101.15 kPa or 1 atm). The uncertainty in the air flow rate measurement is ±1 % with a full-scale reading up to 40 SLPM. During the experiments, two Rosemount 3051s pressure sensors are used to monitor the test section pressure and the pressure difference across the settling chamber and the test section, respectively. The flow speed is derived from the differential pressure between the settling chamber and the test section. The standard errors of the pressure measurements are around 0.1 kPa for both pressure sensors which yields a maximum error of 0.11 m/s in the result of instantaneous flow speed and a mean error around 0.02 m/s. A Validyne DP-15 pressure transducer with the standard error of 0.1 kPa is used to measure the pressure difference across the cavity surface to determine the $\sigma_C$. Overall, in the present experiments, the maximum uncertainties of the $C_{Qf}$, $C_{Qc}$, $Fr$ and $\sigma_C$ are around 2 %. A Nikon D610 DLSR camera is employed to capture images of the overall geometry of the cavity with a 28-mm focal length lens and the detailed cavity morphology near closure region with a 60 mm lens. The supercavity photos are first transferred to grayscale then corrected by un-distortion algorithms based on the record settings in MATLAB$^{TM}$.

## 3. Results

The variations of $C_{Qf}$ upon $Fr$ for all three cavitators are first compared in the present paper (Fig.4). It is worth noting that for the triangle cavitator, the $C_{Qf}$ results are only plotted up to $Fr$ =11.1 (corresponding to a flow speed of 6 m/s) since the ventilation rate to form the supercavity at higher $Fr$ exceeds the operational range of the mass flow controller. At each $Fr$, the $C_{Qf}$ is measured multiple times to ensure statistical robustness, and the uncertainty corresponding to the $C_{Qf}$ is estimated through the standard deviation of all the measured values. The $C_{Qf} - Fr$ curve for disk cavitator mounted on FFM is also compared with the corresponding BFM case from the previous investigation [2]. As shown in Fig. 4, the $C_{Qf}$ initially increases then decreases with increasing $Fr$ for both disk and cone cavitators. Note that the $C_{Qf}$ for FFM-configured disk cavitator matches closely with that for the backward-facing model (BFM) case. For the triangle cavitator, the trend of $C_{Qf} - Fr$ follows that of the disk cavitator in the range of $= 5.6 \sim 11.1$. As suggested by Karn et al. [2] and further evidenced in the observation of aerated hydrofoil wakes [13], in the low $Fr$ regime ($Fr$ <10), the increasing $Fr$ associated with the increase of flow speed affects adversely the bubble coalescence process required to form a supercavity. Consequently, in the low $Fr$ regime, the $C_{Qf}$ rises with increasing $Fr$. After reaching a critical $Fr$ value (i.e. $Fr = 10$, corresponding to the peak $C_{Qf}$), turbulence-induced bubble breakup becomes dominant with further increase of $Fr$. This process significantly enhances the number density of small bubbles in the wake of the cavitator as observed in [13]. Such enhancement of bubble concentration is in favor of the bubble coalescence and thus lowers the $C_{Qf}$ with further increase of $Fr$.

It is worth noting that the $C_{Qf}$ of the cone cavitator is substantially lower than those of other cases under the same $Fr$. Moreover, the critical $Fr$ for the cone cavitator is shifted to higher value compared to its 2D counterparts. As suggested by Calvert [14], the 3D shape of the cone cavitator could induce a more stable flow separation and thus more confined wake under the same $Fr$ compared to 2D-shaped cavitators. Therefore, under low $Fr$, the number density of bubbles in the wake of a cone cavitator is higher, which favors the coalescence process and the formation of the cavity. However, as $Fr$ increases, in comparison to 2D cavitators, the cone cavitator can slow down the development of turbulence in the wake and the corresponding turbulence-induced bubble breakup, leading to the shift of the critical $Fr$ in $C_{Qf} - Fr$ to higher value. As shown in Fig. 4, with further increase of $Fr$, the discrepancy in $C_{Qf}$ between the cone and disk cavitator eventually diminishes. This trend suggests that the cavitator shape effect on the supercavity formation gradually becomes negligible under high $Fr$, potentially due to the intensified turbulent-induced bubble breakup and increasing dissolved gas in the facility. As for the triangle cavitator, the $C_{Qf}$ is slightly higher than that of disk cavitator. The sharp corners on the triangle cavitator may cause unstable and turbulent wake flow in high $Fr$, which adversely affects the supercavity formation driven by the bubble coalescence.

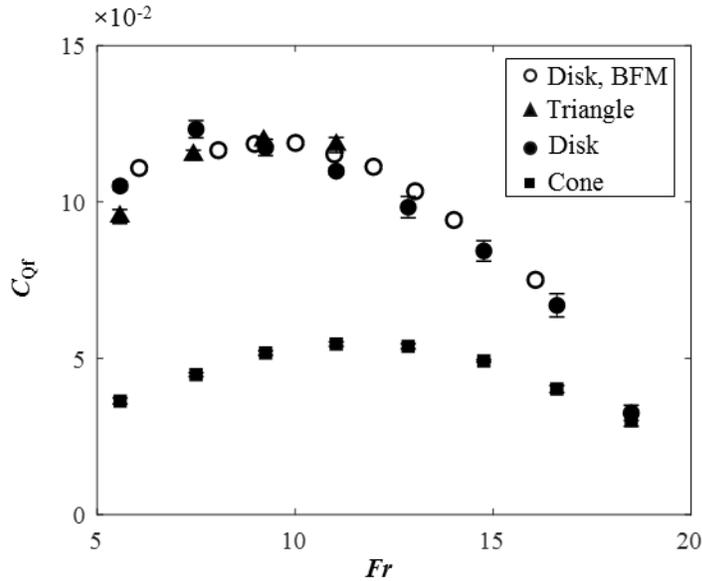

Figure 4: The cavity formation ventilation coefficient $C_{Qf}$ under different $Fr$ for the disk, triangle, and cone cavitators. For the disk cavitator, the data for the forward facing model (FFM) from our experiment is compared with the one for the backward facing model (BFM) from [2].

Fig. 5 depicts the variation of $C_{Qc}$ upon $Fr$ for cavitators with different shapes and mounting struts. The results suggest a much lower $C_{Qc}$ compared to $C_{Qf}$ under the same $Fr$ for each case which has been reported in the previous investigations [2-4]. However, with increasing $Fr$ in the current investigation, the $C_{Qc} - Fr$ curve does not exhibit the same trend, i.e., monotonic decreasing then plateauing, as in [2]. Particularly, the $C_{Qc}$ shows a clear discrepancy between the BFM-configured and FFM-configured disk cavitators. Comparing BFM and FFM of disk cavitator, the FFM case shows a lower $C_{Qc}$ for $Fr < 10$. We attribute such difference to the presence of the ventilation pipe inside the cavity.

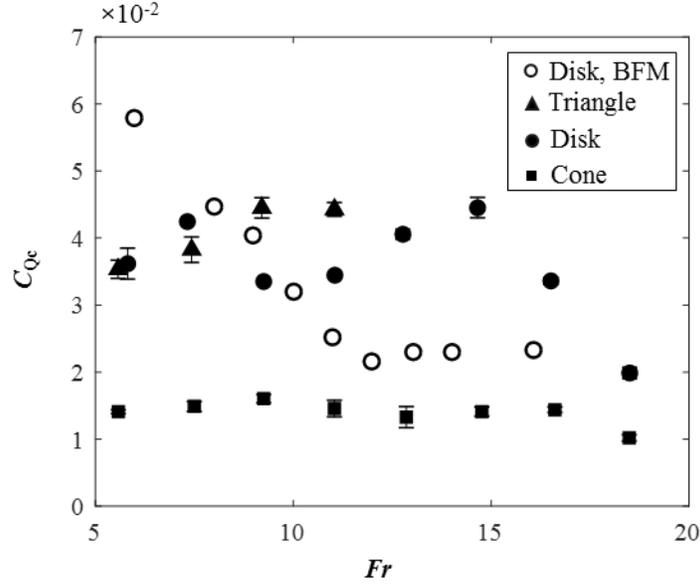

Figure 5: The cavity collapse ventilation coefficient $C_{Qc}$ under different $Fr$ for the disk, triangle cavitators as well as the case of cone cavitator. For the disk cavitator, the data for the forward facing model (FFM) from our experiment is compared with the one for the backward facing model (BFM) from [2].

In the FFM case, the presence of the ventilation pipe allows the supercavity to cling to solid wall near the closure region of the cavity (Fig. 6a), as observed in the hydrofoil supercavitation [15-16] and the ventilated partial cavity generated by a backward facing step cavitator [17]. According to Mäkiharju et al. [17], for low $Fr$ (associated with low tunnel speed $U$), the effect of surface tension is dominant near the cavity closure due to the strong interface curvature and corresponding small Weber number (i.e., $We = \rho_W U^2 l/\sigma$, in which $l$ is a length scale characterizing the local curvature at the closure and $\sigma$ the surface tension coefficient). Such effect of surface tension can suppress the breakup of gas pockets at the closure due to interface instability, leading to a reduction of the ventilation demand to sustain the cavity. However, with increasing $Fr$, the surface tension dominance gradually diminishes with the corresponding increase of $We$ and strong re-entrant water jet emerges at the closure as illustrated in Fig. 6(b). Specifically, as those reported in [15, 17-19], the re-entrant jet forms near the closure region of a cavity due to strong adverse pressure gradient, which can be further enhanced with the presence of a solid wall. In our case, the re-entrant water jet first develops within the gap between the cavity and the ventilation pipe. It then gushes upstream, causing the collapse of the cavity, similar to the cases observed in the cavitation over hydrofoils [15, 18-19] and behind a backward facing step [17]. Note that although the re-entrant jet is also observed in BFM case under low $C_Q$ [4], the presence of solid wall in FFM leads to the formation of stronger re-entrant jet, which promotes the cavity collapse and ultimately yields an increase of $C_{Qc}$ for high $Fr$. According to this mechanism, it is conceivable that the momentum of re-entrant jet should be approximately balanced by that of air ventilation under $C_{Qc}$. To assess this hypothesis, the estimated of re-entrant jet momentum is provided later in the Results Section following a comparison of cavity geometry under $C_{Qc}$ for different cavitator shapes. It is also worth commenting on that the effect of ventilation pipe on the flow near the closure is also manifested by the variation of closure patterns between FFM and BFM cases. As shown in Fig.7, the ventilation pipe induces the formation of re-entrant jet leading to that the FFM-generated supercavity exhibits hybrid twin vortex and re-entrant jet closure in contrast to the exclusive twin vortex of the BFM cases under the same flow conditions [4].

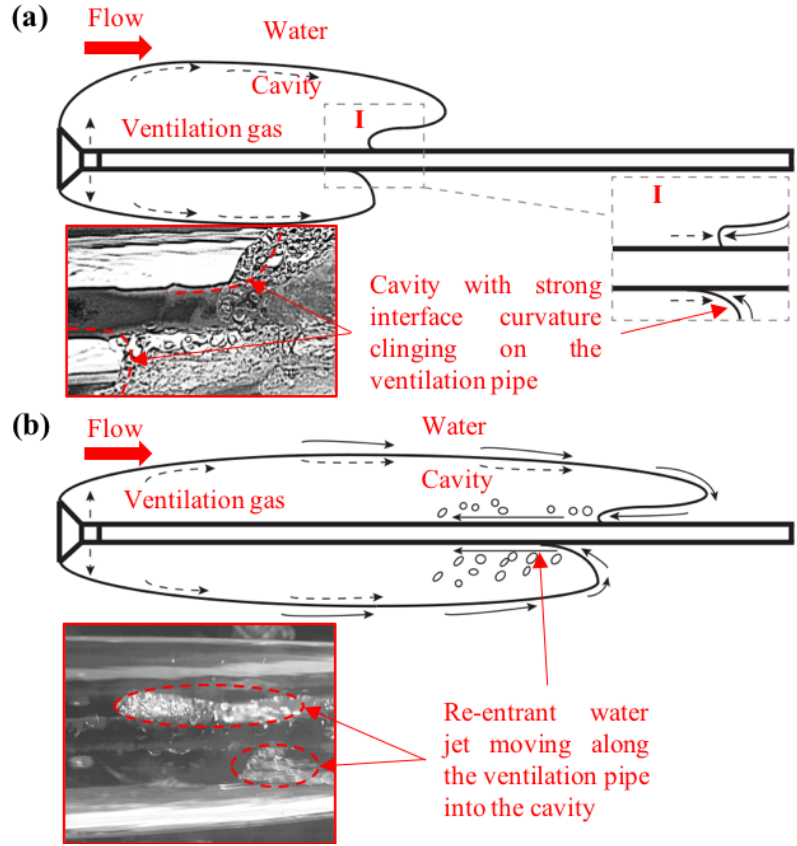

Figure 6: Schematics showing the interaction of ventilated supercavity with the ventilation pipe under (a) low $Fr$ and (b) high $Fr$ number conditions. The inset image in (a) shows the cavity surface near the closure (outlines marked by the red dashed curves) clings onto the ventilation pipe, and that in (b) evidences the presence of re-entrant jet (marked by the red dashed ellipses) in the gap between the cavity and the ventilation pipe under high $Fr$. The black solid and dashed arrows in (a) and (b) denote the direction of water and gas flows, respectively. The inset figure in (b) is a snapshot of a cavity video from the bottom-view high speed imaging to show the re-entrant jet moving upstream of cavity. Such video is used to determine re-entrant jet velocity for estimating jet momentum in the present study.

In addition, comparing the $C_{Qc}$ for different cavitator shapes shown in Fig. 5, the cone cavitator yields the lowest value within the same range of $Fr$. Moreover, the $C_{Qc}$ of the cone cavitator is significantly less dependent on $Fr$ in comparison to the 2D-shaped ones. These trends can be explained through the pressure distribution and the geometry of the supercavity formed by different cavitators. Specifically, the Fig. 8 provides the non-dimensionalized cavity pressure (i.e. cavitation number $\sigma_C$) for different cavitators. As it shows, the $\sigma_C$ for the cone-generated supercavity is consistently lower than those for 2D cavitators in the range of $Fr$ investigated in our experiments, while the values of $\sigma_C$ differ little between the disk and the triangle cavitators. These pressure measurements demonstrate a substantially smaller pressure loss and more confined wake region associated with the cone cavitator, leading to the lowest $C_{Qc}$ of all the cavitator shapes. Moreover, the results also suggest that the more confined wake from the cone cavitator could result in a supercavity with smaller dimension.

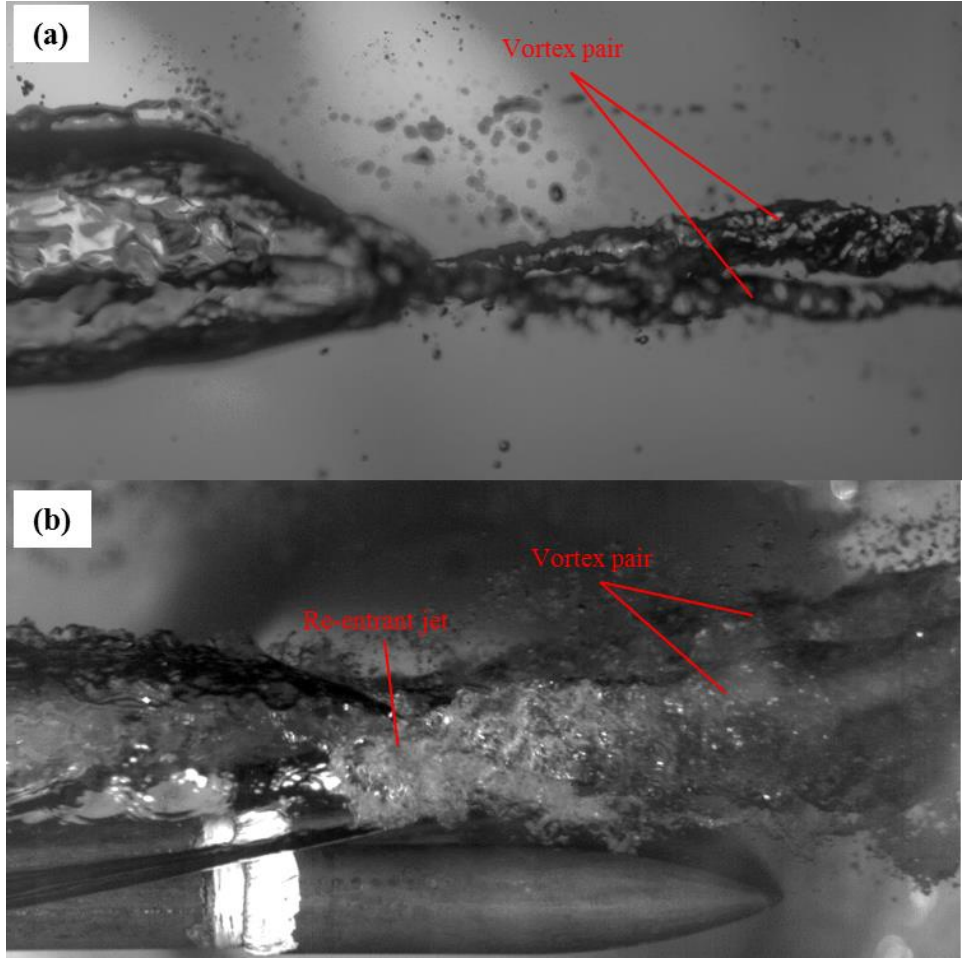

Figure 7: Images showing the ventilated supercavity with (a) twin vortex closure generated by BFM and (b) hybrid twin-vortex and re-entrant jet closure produced using FFM under $d_c = 30.0$ mm, $Fr = 11.1$ and $C_Q = 0.12$.

Subsequently, the geometry of supercavity under $C_{Qc}$ including the overall shape, the cavity maximum diameter ($D_{max}$) and half length ($L_{1/2}$), are investigated for different cavitator shapes to further elucidate the trend of $C_{Qc}$ mentioned above. Such information will be further employed latter to assess the re-entrant jet mechanism that influences the $C_{Qc}$ for disk and cone cavitators. Particularly, according to [3], the $L_{1/2}$ is defined as the distance between the cavitator plane and the location of $D_{max}$. Note that for non-axisymmetric body such as triangle cavitator, the $D_{max}$ obtained from the 2D projection depends on the orientation of the triangle with respect to the projected plane. Therefore, in our experiments, the triangle cavitator is always installed with one of its edges parallel to the bottom window to ensure the consistency of its $D_{max}$ measurement under different $C_{Qc}$ and $Fr$. The difference in the overall geometry of the cavity at limiting ventilation rates (i.e., $C_{Qc}$) among different cavitators is first displayed using sample images recorded under the same $Fr$. As shown in Fig. 9, the cavity generated by the cone and the disk shows a smoother surface as opposed to that of the triangle case whose cavity exhibits distinct contour lines originated from the sharp corners of the cavitator. We suggest that the sharp corners of a triangular cavitator can induced additional flow separation and instability, which perturbs the cavity surface and bubble coalescence. Such perturbation promotes the collapse of a cavity, leading to a slight increase in $C_{Qc}$ compared to the disk cavitator under the same $Fr$ (Fig. 5).

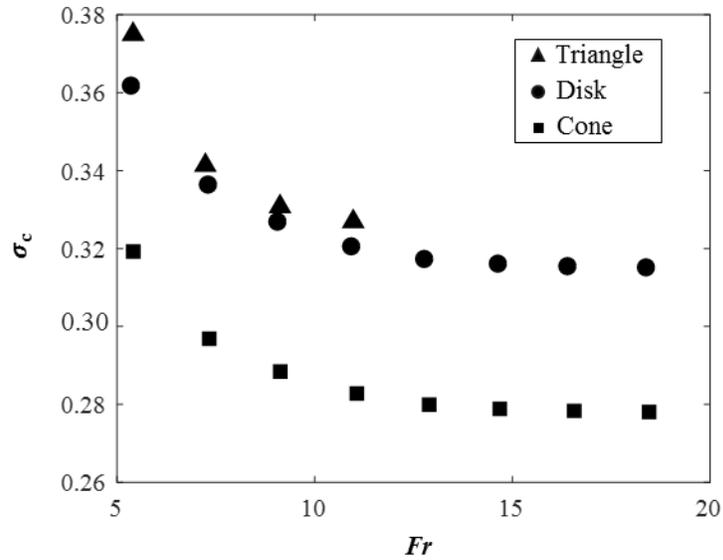

Figure 8: The cavitation number $\sigma_c$ for the triangle, disk and cone cavitator over the range of $Fr$ in our experiments Note that the $\sigma_c$ is independent of $C_Q$ once a clear supercavity is formed [3, 5] and the $\sigma_c$ presented here is obtained at a ventilation coefficient above $C_{Qc}$.

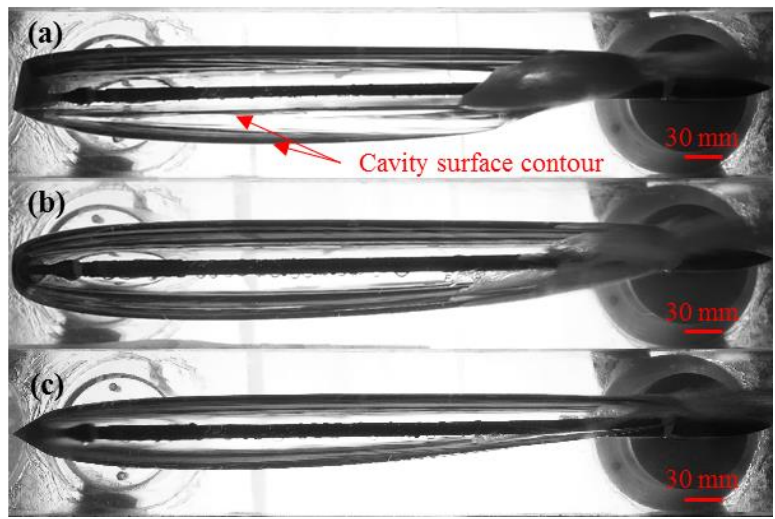

Figure 9: Images of ventilated supercavities formed by (a) the triangle, (b) disk, and (c) cone cavitators at $Fr = 11.1$ under their respective $C_{Qc}$. Contour lines on the surface of the cavity generated by the triangle cavitator are marked by the red arrows.

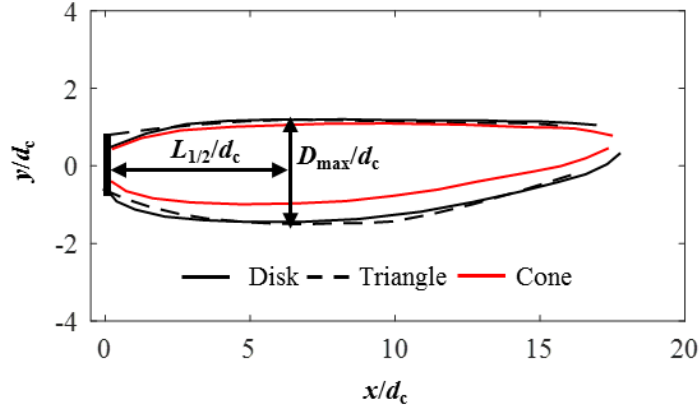

Figure 10: Supercavity geometric profiles extracted from Fig. 8 with normalized maximum diameter ($D_{max}/d_c$) and half length ($L_{1/2}/d_c$) corresponding to the disk cavitator case annotated in the figure by the black arrows.

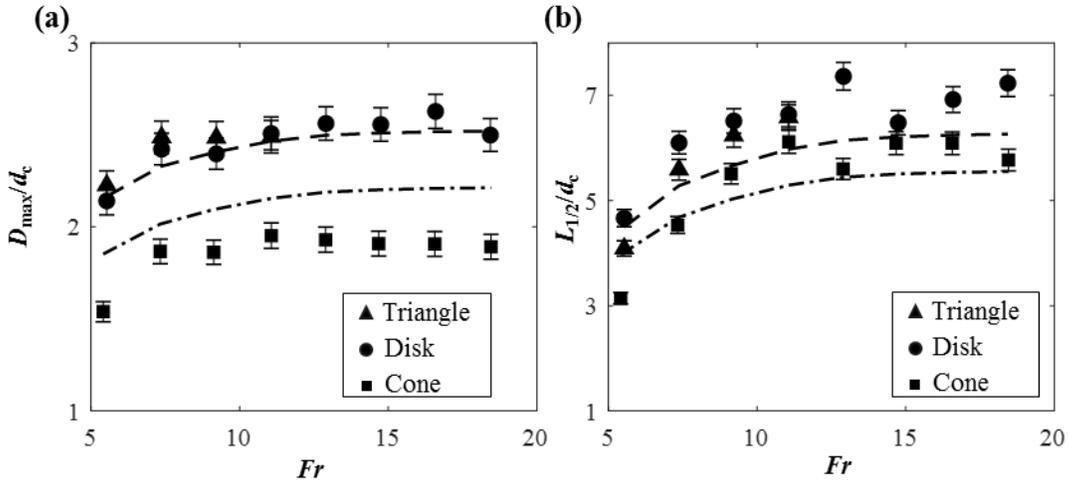

Figure 11: (a) Normalized maximum diameter and (b) half length of the supercavity for the triangle, disk and cone cavitators over the range of $Fr$ in our experiments. All the dimension data is taken from the cavity under their respective collapse ventilation coefficient $C_{Qc}$. The uncertainty of the measurement is about 7.1% from the uncertainty of cavity edge coordinates. The dashed and dot-dashed lines correspond to the semi-empirical formulation from [9] for disk and cone cavitators, respectively.

Additionally, under the same flow conditions, the comparison of the 2D projected supercavity contours from the three cavitators (Fig. 10) shows that the cavities have similar total length, but the cone cavity is significantly narrower in terms of maximum diameter. Furthermore, the difference of the cavity geometry in terms of $D_{max}$ and $L_{1/2}$ under different $Fr$ is quantified (Fig. 11). As shown in the figure, the cone-generated supercavity yields the smallest $D_{max}$ over the range of $Fr$ in our experiments, which is consistent with the pressure measurement results presented in Fig. 8. Additionally, the $D_{max}$ becomes independent of $Fr$ for both disk and cone cases as $Fr$ increases due to the diminishing gravitational effect on the cavity dimension. The measured $D_{max}$ is compared with those calculated from the semi-empirical equation from [9], i.e. equation (1) shown below, for disk and cone cases, respectively. This equation is derived using the potential flow theories with assumption of constant cavity pressure.

$$D_{max}/d_C = \sqrt{C_{x0}(1+\sigma_\infty)/0.96\sigma_\infty} \qquad (1)$$

In the above equation, the $\sigma_\infty$ is equivalent unbounded cavitation number corresponding to each $\sigma_C$ calculated from $\sigma_\infty = (2\sigma_C^2 - \sigma_{min}^2)/2\sigma_C$, where $\sigma_{min}$ is the minimum cavitation number achievable in a closed-wall water tunnel based on cavitator blockage ratios ($\sigma_{min}$ =0.32 in our case) [3]. The $C_{x0}$ is the drag coefficient of the cavitators under $\sigma_\infty = 0$. In the present paper, $C_{x0}$ equals 0.82 for 2D blunt shape cavitators and 0.51 for cone cavitator corresponding to a cone angle of 62°. The semi-empirical equation captures the general trend of the variation of $D_{max}$ upon $Fr$, but over-predicts the values of $D_{max}$ for the cone cavity throughout the range of $Fr$ investigated in our experiments. We suggest that such discrepancy may be caused by the fact that the $\sigma_{min}$ in the formula is only determined by the blockage ratio but does not consider the effect of cavitator shape. As for the $L_{1/2}$, the cavities generated from different cavitators share the same increasing trend upon on the increase of $Fr$ with the cone cavity being slightly shorter than those from the 2D cavitators. The $L_{1/2}$ is also compared with the semi-empirical relation of $L_{1/2}$ upon $\sigma_\infty$ for disk and cone cases from [9] as below:

$$L_{1/2}/d_C = \sqrt{C_{x0}(1+\sigma_\infty)}/\sigma_\infty \qquad (2)$$

As shown in Fig. 11, the equation (2) underestimates the $L_{1/2}$ for all the cases. Such underestimate can be primarily attributed to the variation of test section pressure along the cavity span due to the cavity-induced blockage and friction loss, which is not fully addressed by the semi-empirical formulation based on potential theory.

Using supercavity dimension, the momentum of re-entrant jet ($\dot{M}_W$) can be estimated based on empirical formulation and experiment data to compare with the momentum of ventilation air jet ($\dot{M}_A$) used to sustain the cavity. The $\dot{M}_W$ is estimated using:

$$\dot{M}_W = A_{jet}\rho_W U_{jet}^2 \qquad (3)$$

In the above equation, $A_{jet}$ is the re-entrant jet cross-section area and $U_{jet}$ is the re-entrant jet velocity. The estimation of $U_{jet}$ is from the extent of re-entrant jet into the cavity observed from bottom view high-speed imaging. The $U_{jet}$ is around 10% of $U$ and the disk-generated cavity has consistently higher $U_{jet}$ compared to cone case. According to [15], the dimension of the re-entrant jet developed near closure region for 2D hydrofoil case is expressed as below:

$$\delta = \delta^* + T/(2\rho U_C^2) \qquad (4)$$

where $\delta^*$ is the re-entrant jet thickness without adverse pressure gradient which is typically very small, $T$ is a tangential force term applied to model the adverse pressure gradient over the span of the cavity and $U_C$ is the water flow speed at the gas-liquid interface. Note that $T$ can be calculated as $T = (P_{down} - P_c)\delta_{cav}$, where $P_{down}$ is the pressure at the downstream side of the cavity and $\delta_{cav}$ is the cavity maximum thickness in the 2D case. For the 3D cavity in our case, we suggest that the maximum cavity cross-section area can be used to estimate this tangential force term, i.e., $T = (P_{down} - P_c)A_C$, and $A_{jet}$ is directly calculated from (5):

$$A_{jet} = A^* + T/(2\rho U_C^2) \qquad (5)$$

Similar to [15], in the present case, $A^*$ is the re-entrant jet cross-section area without adverse pressure gradient with a magnitude of $10^{-3}$ mm² (<1% of $A_{jet}$) and is negligible during the calculation according to [15]. Note that $A_C$ is calculated by assuming an annular shape cavity with an outer diameter of $D_{max}$ and an inner diameter same to the ventilation pipe diameter. According to Karn et al. [4], the pressure difference across the cavity surface at cavity downstream ($P_{\text{down}} - P_c$) is estimated as $1/2\,\rho U[\sigma_C(1 - C_{x0}B^2) - C_{x0}B^2]$, where $B$ is the blockage ratio. For the ventilated supercavity in a closed-wall facility, the $U_C$ can be calculated as $U_C/U = \sqrt{1 + \sigma_C}$ according to [20]. As for the ventilation air, the $\dot{M}_A$ can be calculated by the following:

$$\dot{M}_A = \dot{m}_A U_A \tag{5}$$

In the above equation, $\dot{m}_A$ is the mass flow rate of ventilation air and $U_A$ is the flow speed estimated based on size of ventilation holes on the model and cavity internal pressure $P_c$ [21]. Comparison of $\dot{M}_A$ to $\dot{M}_W$ for disk and cone cavitators under their respective $C_{Qc}$ is shown in Fig. 12. For both cases, $\dot{M}_A$ and $\dot{M}_W$ have a reasonable matching at relatively low $Fr$ in our experiments, supporting the re-entrant jet mechanism that dictates cavity collapse and the value of $C_{Qc}$ for cavitators of different shapes. However, the discrepancy between $\dot{M}_A$ and $\dot{M}_W$ enlarges with increasing $Fr$. Particularly, the $\dot{M}_A$ starts to decrease once $Fr$ reaches a critical value despite continuous rising of $\dot{M}_W$ with increasing $Fr$. Note that although it is difficulty to experimentally determine $A_{\text{jet}}$, this increasing trend of $\dot{M}_W$ can be substantiated by the stronger re-entrant jet moving further upstream into the cavity observed from the high-speed videos. We suggest that the discrepancy of $\dot{M}_A$ to $\dot{M}_W$ under high $Fr$ is primarily due to the emergence of localized natural cavitation with increasing $Fr$, which reduces ventilation demand to sustain the cavity. As the cavity transitions to natural supercavitation state with further increasing $Fr$, it is expected that $C_{Qc}$ will gradually reduce to zero [5]. Regarding the cavitator shape, $\dot{M}_W$ for disk-cavitator is shown to be consistently larger than that of the cone case, evidenced from the significantly higher $U_{\text{jet}}$ in the disk case. Our analysis shows that the difference of $\dot{M}_W$ between these two cases is largely (>80%) contributed by the difference in $U_{\text{jet}}$. It is also worth noting that the critical $Fr$ where $\dot{M}_A$ starts to decline shifts to higher value for the cone cavitator case, which is likely due to the delayed inception of localized natural cavitation caused by the smaller pressure loss associated with the cone cavitator (Fig. 8).

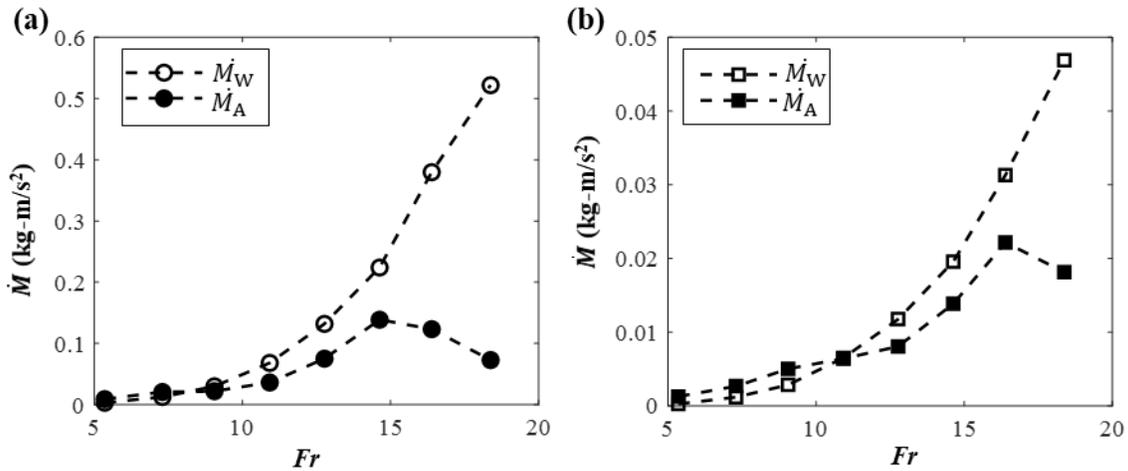

Figure 12: Comparison of momentum of water jet and ventilation air under $C_{Qc}$ for (a) disk case and (b) cone case, respectively.

## 4. Summary and Discussion

We conducted an experimental investigation of the effect of mounting strut and cavitator shape on the formation ($C_{Qf}$) and collapse ventilation coefficients ($C_{Qc}$) of a supercavity under different Froude numbers ($Fr$). Three cavitators including 2D (i.e. triangle and disk) and 3D (i.e. cone-shaped) cavitators of the same frontal area, mounted on a forward-facing model (FFM), are employed. The results are compared with those from a disk cavitator of the same size mounted on a backward-facing model (BFM), in which the cavity is free from the influence of the solid body (i.e., ventilation pipe) within the cavity. Similar to the case of BFM disk cavitator, $C_{Qf}$ initially increases with increasing $Fr$ then decreases after reaching a critical $Fr$ for both FFM disk and cone cavitators. However, the cone cavity yields a substantially lower $C_{Qf}$ and a critical $Fr$ shifting to higher value in comparison to 2D cavitator cases, which may be attributed to the more stable flow separation and confined wake behind a cone cavitator. Accordingly, the cavity generated from a triangle cavitator yields a slightly higher $C_{Qf}$ than that of disk, likely due to more unstable flow separation associated with the sharp corners of a triangle. The $C_{Qc} - Fr$ trend of the FFM-configured cavitator deviates from that of the BFM due to the difference in mounting configurations. Specifically, at low $Fr$, the FFM disk cavity clings on the ventilation pipe near its closure region resulting in a lower $C_{Qc}$ compared to its BFM counterpart owing to the dominant effect of surface tension. However, as $Fr$ increases, the presence of solid surface inside the cavity in FFM enhances the adverse pressure gradient near cavity closure and leads to the formation of strong re-entrant jet along the surface, which promotes the cavity collapse and yields an increase of $C_{Qc}$ compared to that of the corresponding BFM. Regarding the cavitator shapes, the cone cavitator has the smallest $C_{Qc}$ among all the cavitators across the range of $Fr$ in our experiments. To elucidate the trends of $C_{Qc}$ upon changing $Fr$ and cavitator shape, the geometry of supercavity under $C_{Qc}$ including its overall shape, the cavity maximum diameter ($D_{max}$) and half length ($L_{1/2}$) are also investigated. Both $D_{max}$ and $L_{1/2}$ show an increasing then plateauing trend upon increasing $Fr$ across different FFM cavitators despite the noticeable smaller values for the 3D cone cavitator. Subsequently, such cavity geometric information and cavity pressure measurements in conjunction with high speed imaging of re-entrant jet are employed to estimate the re-entrant jet momentum ($\dot{M}_W$) under different $Fr$ for disk and cone cavitators. The estimated re-entrant jet momentum shows reasonable match with the ventilation air momentum ($\dot{M}_A$) under $C_{Qc}$ in lower $Fr$ for both cavitator cases with the disk cavitator case yielding significantly stronger re-entrant jet, providing support to the re-entrant jet mechanism governing on the cavity collapse. However, when $Fr$ rises above certain critical value, the difference between $\dot{M}_W$ and $\dot{M}_A$ enlarges and $\dot{M}_W$ starts declining with increasing $Fr$ potentially due to the inception of localized natural cavitation in this range of $Fr$.

Our study sheds some light on the cavitator design and ventilation strategy for a supercavitating vehicle in practice. Specifically, the cone shape cavitator with a lower drag coefficient not only drastically enhances the ability of drag reduction as previously reported [8, 9] but is also favorable for cavity formation and sustenance. In addition, our study suggests the presence of solid body inside a supercavity can potentially lead to change of internal flow and pressure distribution, which can ultimately affect the stability and the sustenance of a cavity under different ventilation and flow conditions. Noteworthily, Wu et al [22] has recently investigated the cavity internal flow for a BFM disk cavitator. Their results demonstrated the importance of internal flow on the air leakage mechanism and the development of a supercavity upon changing flow conditions. However, such experimental investigation has not been conducted on a FFM cavitator which is more relevant to the supercavitating devices used in practice with a solid body inside the cavity. Therefore, it would be of great interest for future research to look into how the internal flow and pressure field inside a ventilated supercavity can be modified by the presence of a solid body. Particularly, understanding such internal flow can help us establish more detailed models for the formation of re-entrant jet which influences significantly on the collapse and the sustenance of a supercavity as discussed in the current study.

## Acknowledgements

This work is supported by the Office of Naval Research (Program Manager, Dr. Thomas Fu) under grant No. N000141612755 and the Undergraduate Research Scholarship program at the University of Minnesota.